\documentclass[aps,pre,twocolumn]{revtex4}
\usepackage{epsfig}

\def\be{\begin{equation}}
\def\ee{\end{equation}}
\def\ba{\begin{array}}
\def\ea{\end{array}}
\def\bea{\begin{eqnarray}}
\def\eea{\end{eqnarray}}

\begin{document}

\def\Fig{/home/damien/Mypapers/Hier/Fig}

\title{Large scale numerical simulations of ``ultrametric'' long-range
depinning}

\author{Damien Vandembroucq  and St\'ephane Roux}

\affiliation
{Unit\'e Mixte CNRS/Saint-Gobain ``Surface du Verre et Interfaces''\\
39 Quai Lucien Lefranc, 93303 Aubervilliers cedex, FRANCE}

\begin{abstract}
The depinning of an elastic line interacting with a quenched disorder
is studied for long range interactions, applicable to crack
propagation or wetting. An ultra-metric distance is introduced instead
of the Euclidean distance, allowing for a drastic reduction of the
numerical complexity of the problem. Based on large scale simulations,
two to three orders of magnitude larger than previously considered, we
obtain a very precise determination of critical exponents which are
shown to be indistinguishable from their Euclidean metric
counterparts. Moreover the scaling functions are shown to be
unchanged. The choice of an ultrametric distance thus does not affect
the universality class of the depinning transition and opens the way
to an analytic real space renormalization group approach.
\end{abstract}
\pacs{PACS numbers }
\maketitle

\section{Introduction}

The depinning of an elastic interface in a random environment gives a
common theoretical framework to describe physical phenomena as various
as the advance of a magnetic wall, the propagation of a fracture front
or the wetting of a disordered surface (see {\it e.g.}
Ref. \cite{Leschhorn-AnnPhys97,Kardar-PR98,Fisher-PR98} for a recent
review). The richness of the physics encountered in these different
situations results from the same feature. The disorder of the
environment which tends to anchor the interface competes with an
elastic-like term which tends to keep the interface smooth. The tuning
of an external driving force allows to go through a critical
transition. Below threshold the interface can only advance over a
finite distance before stopping in a blocked conformation. Above
threshold the interface can move freely and acquire a finite
velocity. At threshold the system is characterized by a set of
universal critical exponents.

In the case of over-damped dynamics the motion can be described by the
following stochastic equation:

\be \mu \partial_t h(x,t) = F_{ext}(t) + F_{el}(x,t) +
\gamma[x,h(x,t)] \ee where $h(x,t)$ denotes the position of the front,
$F_{ext}(t)$ the external driving force, $F_{el}(x,t)$ the elastic
term due to the distortion of the front and $ \gamma[x,h(x,t)]$ the
frozen disorder.  Depending on the physical phenomenon considered the
elastic term can take different forms. In the case of magnetic walls
or of wetting in a Hele-Shaw cell, the interactions to be taken into
account are short ranged and at first order the elastic force can be
estimated by a simple Laplacian term: $F_{el}(x,t)=\nabla^2 h (x,t)$. In
the case of the advance or receding of a triple line on a disordered
substrate \cite{Joanny-JCP84} or of the propagation of fracture front long
range interactions have to be considered. In the latter case elastic
interactions are mediated {\it via} the bulk along the whole front. A
first order perturbation analysis for  the front roughness
gives\cite{GaoRice-JAM89}:

\be 
 F_{el}(x,t) = \frac{1}{\pi} \int dx' \frac{h(x,t)-h(x',t)}{|x-x'|^2}
\ee

This long range elastic string model has been widely studied over the
last ten years. In particular the interface is shown to exhibit a
self-affine roughness: the width $w$ of the interface scales with the
system size $L$ as $w\propto L^\zeta$.  The different numerical works
\cite{Dong-PRL93,Schmittbuhl-PRL97,Tanguy-PRE98,Rosso-PRE02} performed
give estimations of the roughness exponent $\zeta$ spreading in the
interval $[0.34-0.40]$. The latter exponent and more generally the
universality class of the model depend strongly on the long range
nature of the kernel in Eq. (2). For instance, if $F_{el} \propto
\nabla^2h$, then $\zeta=1.25$.  The most recent results obtained by
Rosso and Krauth \cite{Rosso-PRE02} give a value $\zeta\approx0.39$
significantly larger than the theoretical prediction $\zeta=1/3$
obtained by one-loop calculations of renormalization group
technique\cite{Narayan-PRB93,Ertas-PRE94} and equally smaller than the
recent two-loop estimation\cite{Chauve-PRL01} $\zeta\approx
0.47$. Note however that these values are not consistent with the
results $\zeta\approx0.5-0.6$ obtained experimentally for interfacial
fracture\cite{Schmittbuhl-PRL97} or
wetting\cite{Prevost-PRL99,Moulinet-EJPE02}.  Therefore it is of
uttermost importance to have an accurate determination of these
critical exponents, and thus to be able to study large system sizes.


In the following we present an ultrametric version of an extremal
model of depinning (see Ref. \cite{Tanguy-PRE98} for a detailed study
of the original model with Euclidean metric).  The complexity of the
elementary step of calculation is shown to scale with the system size
$L$ as $\log_2 L$ instead of $L$ in the original model.  This allows
us to perform simulations on systems of size $L=2^{20}\approx 10^6$,
which corresponds to a gain of two to three orders of magnitude
compared with other published works based on the Euclidean
metrics. The universality class of the model is shown to be unchanged.
Beyond the numerical acceleration, this ultrametric model of depinning
may thus also serve as a starting point for a real space
renormalization analysis.

The paper is organized as follows: In a first part we recall the
definition of the original model, we then define the ultrametric
version and present the main features of the new model. In the second
part we give results of simulations performed on large systems and
focus on the critical properties of this ultrametric model of
depinning. We conclude that it lies in the same universality class as
its euclidean version. The details of the numerical implementation of
the algorithm are finally given in an appendix.

\section{Extremal driving of a depinning front}

Various numerical techniques can be used to study the depinning
phenomenon in the vicinity of the critical threshold. Early works used
a direct integration of the equation above {\it via}
Euler\cite{Dong-PRL93} or Runge-Kutta \cite{Zapperi-EPJB00} schemes. 
Recently Rosso and Krauth \cite{Rosso-PRE02} developed  an iterative
algorithm to determine the blocked conformations corresponding to a
given constant forcing. They used periodic boundary conditions both
along the front and in the direction of propagation and the critical
threshold is reached when the last blocked conformation starts moving.

In the case of strong pinning the advance of the front proceeds by
successive local instabilities. This avalanche behavior is
characteristic of the motion of a depinning front. This property can
be exploited to develop an efficient algorithm describing the motion
of the front close to the critical threshold. Instead of driving the
front at a constant external force, it consists of tuning the latter
at the exact value such that one and only one site can depin at a
time. So doing, the sequence of depinning events is preserved. The
implementation of this extremal dynamics which has been used since
1992 in various interface growth models
\cite{Zaitsev-PhysicaA92,Sneppen-PRL92,BakSneppenPRL93,Leschhorn-PRE94,RouxHansen-JP1-94,Paczuski-PRL95}
is straightforward. At each iteration step one needs to identify the
weakest site, to advance it up to the next trap and to update the long
range elastic forces due to the change of front conformation. The
latter operation scales with the size of the front. The great
advantage of this method is that the system remain constantly at the
edge of the critical behavior and it is not necessary to tune the
external driving force (see {\it e.g.} Ref. \cite{Zapperi-BJP00} for a
discussion on the use of extremal dynamics to reach the critical state
and more generally on the link between Self Organized Criticality and
classical critical transitions).


Based on the above described extremal modeling, we now detail the way
to implement the model numerically.  The front is discretized along a
regular horizontal grid; $i \in [0,L-1]$ and $h_i$ are the coordinates
along the front and in the direction of propagation
respectively. Traps of random depths $\gamma_i$ are randomly distributed
along the direction of propagation. The distortion of the front
induces elastic forces $f_i^{el}$ {\it via} a Green function
$G_{ij}=G(r_{ij})$ where $r_{ij}$ is the euclidean distance separating
two sites $i$ and $j$ along the front. In the case of a fracture
front, a discretized version of the elastic redistribution function is
such that \be G_{ij}\propto_{i\ne j} |i-j|^{-2} \;, \quad G_{ii}=-\sum_{i\ne j}
G_{ij}\;.  \ee

 Let us consider a given conformation of the front. For each site
$(i,h_i)$ located in a trap of depth $\gamma_i$, we can define a local
depinning threshold $s_i=\gamma_i-f_i^{el}$: this site depins as soon as
the external driving force $F$ overcomes the threshold $F> s_i$. The
depinning threshold $s(t)$ of the front conformation obtained at
iteration $t$ thus corresponds to the minimal external force to be
exerted so that at least one site of the front can depin: $s(t)=
\min_i s_i$. Finally the critical threshold $s^*$ above which the
front can freely propagate is $s^*=\max_t s(t)$. The basic rule of the
extremal driving consists simply at each iteration step $t$ of simply
tuning the external force at exactly the value of the depinning
threshold of the current front conformation: $F(t)=s(t)$. Once
identified, the extremal site is  advanced up to the next trap,
the elastic forces are  updated to take into account this local
displacement and the new value $s(t+1)$ of the front depinning
threshold is evaluated.  We summarize below the elementary steps of
the algorithm used to run the model and we estimate their complexity
respectively to the size $L$ of the system.

\begin{center}
\begin{tabular}{lll}
{\bf A} &initialization $h_i\leftarrow 0\;, f_i^{el} \leftarrow 0\;,
 \gamma_i\leftarrow rnd $
  &$[L]$ \\
\vspace{4pt}
& $i \in[0-L]\;, \quad L=2^n$ &\\
{\bf B} &identification of the extremal site $i_0$ & $[L]$ \\
\vspace{4pt}
&such that $\gamma_{i_0}-F_{i_0}=\min_i (\gamma_i-f_i^{el})$ &\\
{\bf C} &advance of the extremal site  &$[1]$\\
\vspace{4pt}
&$\delta h_{i_0}\leftarrow rnd \;,  h_{i_0}\leftarrow h_{i_0} +\delta h_{i_0}$ &\\
\vspace{4pt}
{\bf D} &update of the trap depth  $\gamma_{i_0}\leftarrow rnd$ &$[1]$ \\
{\bf E} &update of the elastic forces $f_i^{el} \leftarrow f_i^{el} +
G_{ii_0}\delta h_{i_0}$  &$[L]$ \\
\vspace{4pt}
&where $G_{ij} \propto |i-j|^{-2}$ &\\
{\bf F} &back to step B &
\end{tabular}
\end{center}
where $rnd$ stands for a random number and $a \leftarrow b$ for the
assignment of value $b$ to variable $a$.  Except the first
initialization step, the two limiting steps are the identification of
the extremal site and the update of elastic forces along the front
which both scale linearly with the system size $L$. This sequence of
elementary steps is then iterated $T$ times to obtain statistical
averages of the quantities of interest.

\section{Ultrametric depinning}
We now turn to the presentation of the ultrametric  model. The basic
rules of the extremal model remain identical but the redistribution of
the elastic forces.  Instead of using the natural euclidean distance
along the front we use an ultrametric distance.  The structure of the
algorithm stays roughly similar to the previous one but steps {\bf B}
and {\bf E} are shown to be characterized by a complexity in $\log_2
L$ instead of $L$.

\begin{figure}[t]
\epsfig{file=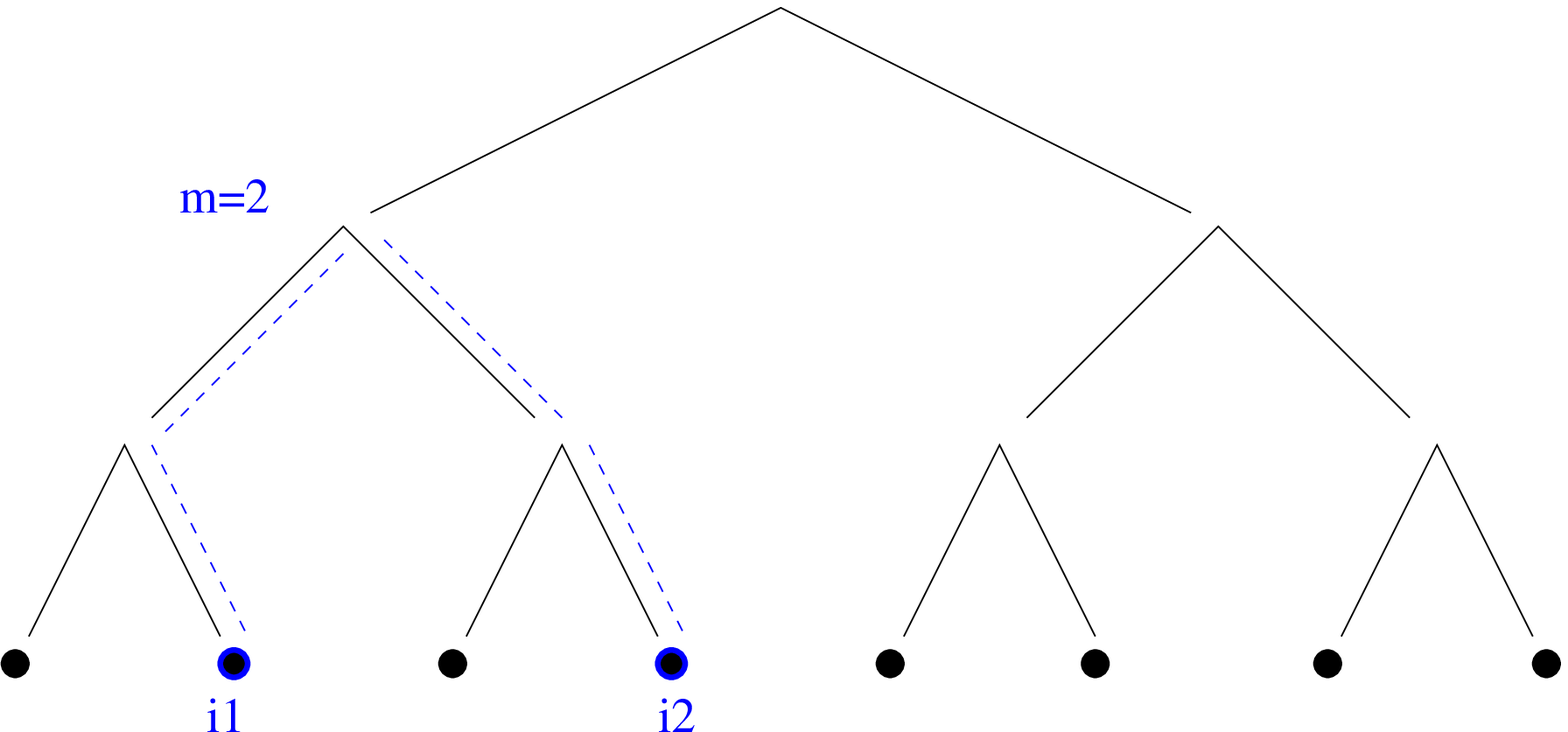,width=0.85\hsize}
\caption{\label{tree_dist} The depth $m$ of the first common ancestor
of sites $i_1$ and $i_2$ is used to compute the ultrametric distance
between points $i_1$ and $i_2$ as $d_u(i_1,i_2)=2^m-1$}
\end{figure}

The most natural structure to be used in the context of a model with
ultrametric distance is a dyadic tree. Let us first build such a tree
whose final leaves are the $L$ sites of the depinning front. As
illustrated on Fig. \ref{tree_dist}, the simplest ultrametric distance
between two sites $i$ and $j$ is the number of branches that composes
the shortest path on the tree between the two sites. This is exactly
twice the depth $m$ of the nearest common ancestor of these two sites.
Different choices of distance are possible based on the tree
structure. In the following we use a definition that preserves the
scaling of the Euclidean distance:

\be
d_u(i,j)=2^m-1
\ee
An important characteristic of this ultrametric distance is its
degeneracy. Namely there is one point at distance $d=2^1-1=1$, two
points at distance $d=2^2-1=3$ and $2^{p-1}$ points at distance
$d=2^p-1$. For a set of $L=2^n$ points one thus counts only $n=\log_2
L$ different values for the distance between two points of the
set. There lies the main advantage of the choice of an ultrametric
distance from the computational complexity point of view. The
expression of the elastic Green function then derives directly from
the definition of the new distance:

\bea
G_{ij}=G\left[d_u(i,j)\right] \propto \frac{1}{d_u(i,j)^2}\\
\nonumber
G_{ii}=-\sum_{m=1}^{\log_2 L} 2^{m-1} G\left[2^m-1\right]
\eea

\begin{figure}[t]
\epsfig{file=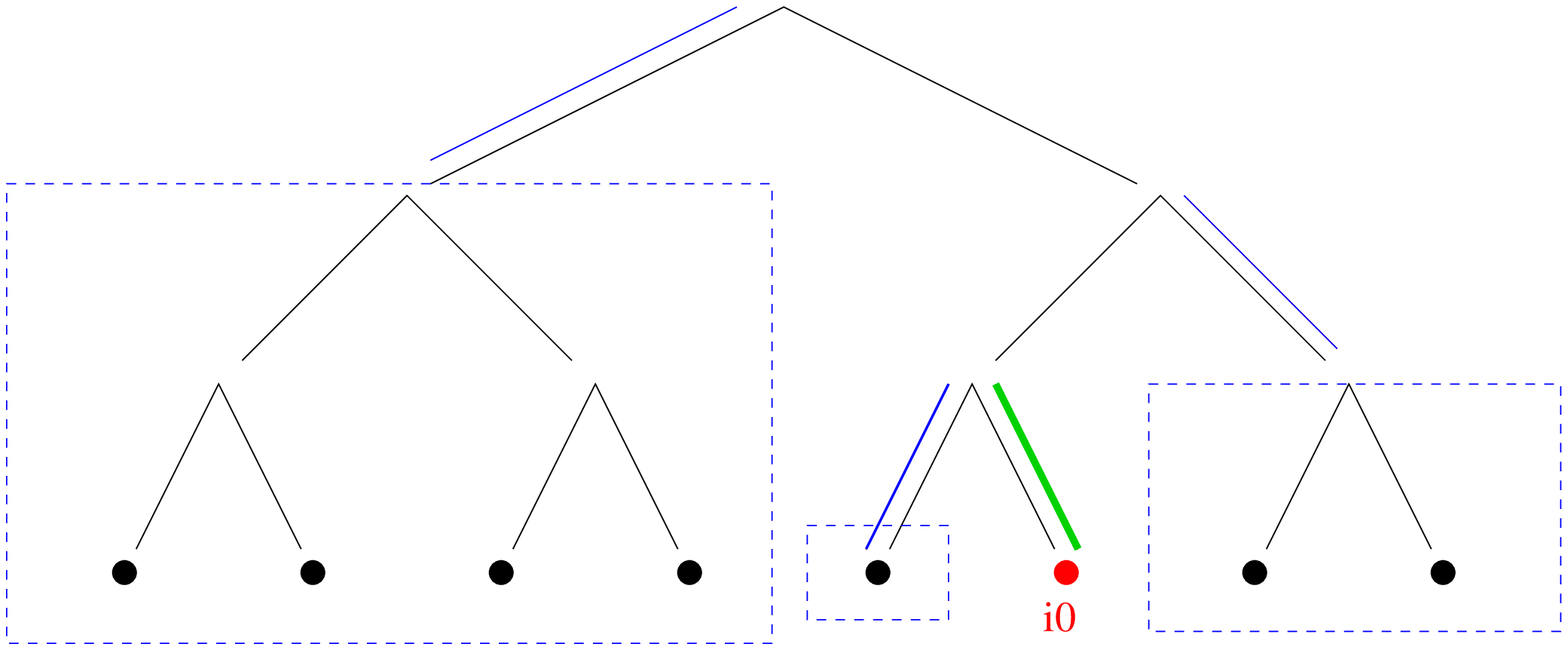,width=0.85\hsize}
\caption{\label{tree_force} 
When redistributing elastic forces after the depinning of the extremal
site $i_0$, all sites located at the same ultrametric distance
$m=d_u(i,i_0)$ belong to a common subtree and receive the same
contribution. This allows an update by block that scales with the
number $\log_2 L$ of these subtrees. Similarly the tree structure
allows to find the next extremal sites in only $\log_2 L$ operations.}
\end{figure}

Using this definition, we can easily accelerate the update of elastic
forces (step E). As illustrated on Fig. \ref{tree_force} instead of
updating $L$ sites, we can update only $\log_2 L$ subtrees
corresponding to sites located at the same ultrametric distance of the
extremal site.  A similar gain can be obtained on  step B for the
determination of the extremal site. Technical details regarding the
numerical implementation of the algorithm are developed in the
appendix. The basic steps of the algorithm then scale with $\log_2 L$
instead of $L$.  The price to pay for this numerical advantage is the
loss of the translational invariance. We show below that it does not
affect the universality class of the model.

The simulations have been performed on systems of sizes up to
$L=2^{20}$. The numerical runs were performed over a large duration
$T$ respective to the natural correlation time of the system
$\tau_L\propto L^z$. In the case of the largest system $L=2^{20}$ we
used $T=2.5 10^{10} \approx 100 \tau_L$.

\section{Characterization of the critical state: numerical results}

The propagation of depinning fronts at the critical threshold exhibits
a rich phenomenology. The front presents a self-affine roughness: its
width $w(\Delta x)$ measured over a distance $\Delta x$ scales as
$w(\Delta x)\propto \Delta x^\zeta$, where $\zeta$ is the roughness
exponent. 

The dynamics of depinning is characterized by an avalanche
behavior. In the framework of an extremal dynamics this can be
described by the distribution of the distances $r$ between two
successive depinning sites: $P(r)\propto r^{-a}$. Generalizing this
distribution for sites corresponding to depinning events separated by a
given time lag allows to obtain in addition the dynamic exponent $z$
which characterizes the spreading of the avalanches. Namely the
lateral extension $\xi$ of an avalanche of duration $\Delta t$ scales as
$\xi\propto \Delta t^{1/z}$.

Another quantity of interest is the external force $s$ needed to depin
a given conformation of the front. It can be shown
\cite{SVR-IJMPC02,VR-preprint03} that the distribution ${\cal Q}(s)$
of these front depinning forces exhibits a singular behavior close to
the critical threshold $s^*$: ${\cal Q}(s) \propto (s^*-s)^\mu$.

In the following we present simulations of ultrametric depinning
performed on large systems (up to $L=2^{20}$). We recover all critical
features described above with exponents numerically indistinguishable
from their counterparts in the Euclidean version of the model.

\begin{figure}[t]
\begin{center}
\epsfig{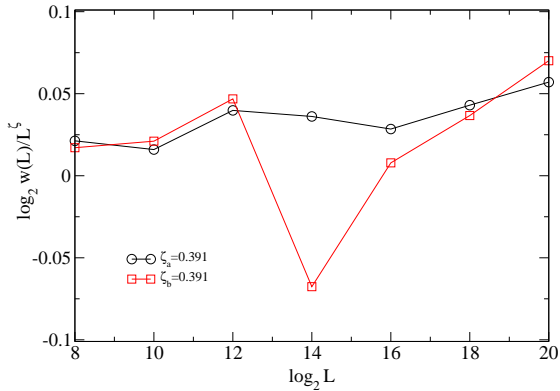}
\end{center}
\caption{\label{width} Width $w(L)$ of the front (standard deviation
of the height distribution) for growing lateral sizes $L$ after
normalization by a power law of exponent $\zeta$. The simulations have
been run from over $100\times 10^6$ iterations steps for $L=2^8$ up to
$25\times 10^9$ for $L=2^{20}$. The circle and the square symbols
correspond respectively to a uniform and a Gaussian distribution of
the trap depth. }
\end{figure}

\subsection{Self-affine roughness}

Various statistical roughness estimators can be used to characterize
the self-affine properties of a rough front. Consider for example the
standard deviation $\sigma(\Delta x)$ of the height differences
between points separated by a distance $\Delta x$. A self-affine front
obeys $\sigma(\Delta x)\propto \Delta x^\zeta$. Similarly the width
$w(L)$ of a front of length $L$ ({\it i.e.} the standard deviation of
the height distribution along the front) scales as
$w(L)\propto L^\zeta$. Fourier or wavelet transforms are also of standard
use. 

In the context of this study we can also design a ``wavelet like''
roughness estimator which exploits the natural tree structure
associated to the ultrametric distance.
At the level $\ell=n=\log_2 L$ we define $\omega^2(n)$ as the variance of the height
difference between nearest neighbors:

 \be \omega^2(n)= 
\left\langle \left( h_i -h_j \right)^2 \right\rangle_{d_u(i,j)=2^n-1}
\ee
 At the
upper level the height of a node is thus simply defined as the
arithmetic average of its two ancestors:
$h^{(\ell-1)}(i)=[h^{(\ell)}(2i)-h^{(\ell)}(2i+1)]/2$ and the
corresponding variance $w(m)$ is computed. This sequence is iterated up
to the root of the tree. At each level $m$ corresponds an ultrametric
distance $m=n-\ell+1$ and we have 
\be
\omega^2(m) \propto 2^{2\zeta(m-1)} \;,\quad {\rm or}\quad \omega(m) 
\propto d_u(m)^{\zeta}
\ee

\begin{figure}
\begin{center}
\epsfig{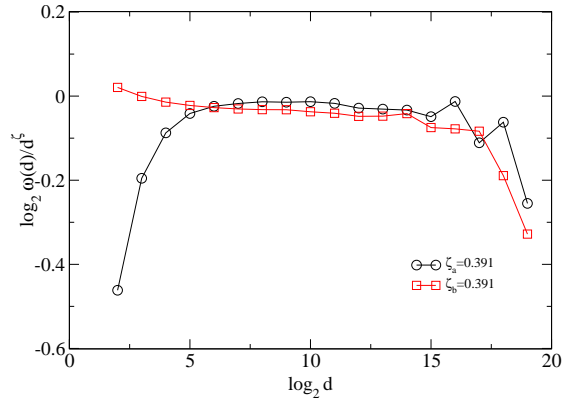}
\caption{\label{wave_hier} Wavelet roughness estimator $\omega(m)$
against the ultrametric distance $d_u(m)=2^m-1$.  The simulations have been
performed on a system of size $L=2^{20}$ run over $25\times 10^9$
iteration steps. The circle and the square symbols correspond
respectively to a uniform and a Gaussian distribution of the trap
depth.}
\end{center}

\end{figure}

We present now numerical results obtained for these various roughness
estimators. The simulations have been performed on systems of sizes up
to $L=2^{20}$. In Fig. \ref{width} and \ref{wave_hier} we show the
scaling behaviors obtained for the wavelet roughness estimator
$\omega(d_u)$ and the width $w(L)$ of the interface. We obtain perfect
power law behaviors over six decades and we only show here in
logarithmic scale the residuals after normalization by a power law of
exponent $\zeta$. This procedure is a very sensitive way of
detecting deviations from a power law. Note that previous published
works deal with $\log_2 L \le 10$.

We present simulations performed with two kinds of distributions for
the trap depths, uniform and Gaussian (respectively denoted by the
subscripts $a$ and $b$ in the figures). 

An estimation of the roughness exponent can be extracted from each
individual set of data. Note that the fluctuations of this estimate
due to the choice of the nature of disorder or the roughness estimator
are larger than the deviation form the power law behavior itself.  All
results are presented here with the central value $\zeta=0.391$. This
value slightly underestimates the results obtained from the width
estimator and slightly overestimates the results obtained from the
wavelet estimator. A conservative estimate of the roughness exponent
thus appears to be:
 \be \zeta=0.391 \pm0.005 \ee

\subsection{Avalanche behavior \--- Dynamic exponent}

The avalanche behavior is characteristic of the dynamics of the front
close to threshold forcing. Although an extremal driving does not
allow to recover the real dynamics of the front, avalanches associated
to a given level of the driving force $F$ can be reconstructed from
the history of the extremal force signal $s(t)$. An avalanche thus
consists of a continuous series of depinning events such that
$s(t)<F$. Instead of reconstructing these avalanches it is classical
in the framework of extremal models \cite{Furuberg-PRL98} to work
directly from the extremal force signal. Let us consider a time lag
$\Delta t$ (a number of iterations), we introduce the distance $\Delta
x$ along the front between the sites depinning at $t$ and $t+\Delta t$
respectively. It appears \cite{Tanguy-PRE98} that the distributions of
these distances $\Delta x$ at fixed time lad $\Delta t$ can be
rescaled on a universal form:

\be P(\Delta x; \Delta t) = \frac{1}{\Delta x^a}
\psi\left(\frac{\Delta x}{\Delta t^{1/z}}\right) \ee where $\psi(u)
\propto u^a$ for $u \ll 1$ and $\psi(u) \approx \mathrm{cste}$ for $u
\gg 1$. The exponent $a$ is well approximated by the exponent of
elastic kernel of Eq. (2) $a\approx 2$ and the dynamic exponent $z$ can be
related to the roughness exponent: if a sequence of $\Delta t$
depinning events spreads over a distance $\Delta x$ along the front,
the knowledge of the roughness of the front over $\Delta x$ leads to
$\Delta t \approx \Delta x \Delta x^\zeta$ thus
$z=1+\zeta$\cite{Tanguy-PRE98,Krishnamurthy-EPJB00}.

\begin{figure}
\begin{center}
\epsfig{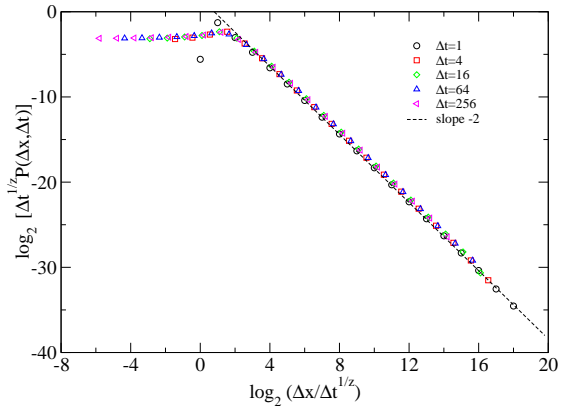}
\end{center}
\caption{\label{furuberg} After rescaling the distribution of
ultrametric distances $d_u$ between sites depinning at time $t$ and
$t+\Delta t$ collapse on a unique master curve. The dynamic exponent
used for the rescaling is $z=1.39$.}
\end{figure}

This scaling is recovered in the framework of ultrametric depinning,
where we measured $P(d_u,\Delta t)$. We check on Fig. \ref{furuberg}
that after rescaling all distributions collapse on a unique master
curve. The rescaling was obtained with the value $z=1.39$, for large
arguments the behavior of $\psi$ is well approximated by a power law
of exponent $a=2$.

\section{Scaling functions}

In addition to critical exponents, the pinned state is also
 characterized by scaling functions\cite{Racz-Houches02}.  In the
 specific case of interface dynamics, this means that the critical
 properties of the front are described by a universal exponent and a
 universal function describing the fluctuations of the (rescaled)
 width of the interface. This property has been evidenced for various
 growth models (Edwards-Wilkinson,
 Kardar-Parisi-Zhang...)\cite{Racz-Houches02} and has been recently
 applied to the case of depinning interfaces \cite{Rosso-PRE03}. Note
 that in the latter case, beyond the interface width, the technique
 can be used to characterize other fluctuating quantities. In
 particular, the distribution of the depinning threshold of a finite
 elastic line under extremal driving can be shown to be
 universal\cite{SVR-IJMPC02,VSR-preprint03}. In the present study we
 show that the choice of an ultrametric distance does not affect these
 universal distributions. More precisely the fluctuations of interface
 width and depinning threshold are shown to be described by universal
 functions and these functions appear to be very close to or identical
 to their counterparts obtained in the framework of a depinning model
 with Euclidean distance.

These statistical distributions are however sensitive to the boundary
conditions (periodic boundary conditions {\it vs} isolated
system)\cite{Antal-PRE02}. In the following we use periodic boundary
conditions. In the ultrametric case the re-summation over all replicas
induces an additional mean field contribution $1/2L^2$ equally shared
by all $G_{ij}, i\ne j$.

\subsection{Universal width fluctuations}

\begin{figure}
\begin{center}
\epsfig{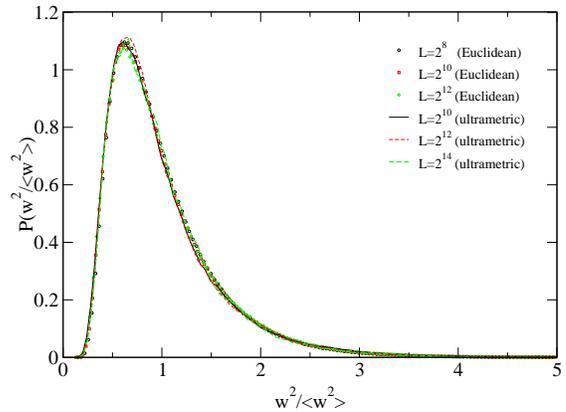}
\end{center}
\caption{\label{distri-width} Distribution of the rescaled width of
the depinning front for sizes $L=2^{10},\; 2^{11}$ and $L=2^{12}$ in
the Euclidean model and $L=2^{10},\; 2^{12}$ and $L=2^{14}$ in the
ultrametric model. All distributions collapse on the same curve.}
\end{figure}

Following Ref. \cite{Rosso-PRE03} we study the distribution of the
rescaled width $w^2 /\langle w^2 \rangle$ where $\langle w^2 \rangle$
is the temporal average of the width of a depinning front of finite
extent $L$. In Fig. \ref{distri-width} we present results obtained for
various system sizes in both cases of Euclidean and ultrametric
distance. We observe that all rescaled distributions collapse onto a
master curve whatever the size of the front and the Euclidean or
hierarchical metric.

\subsection{Universal depinning force fluctuations}

\begin{figure}[t]
\begin{center}
\epsfig{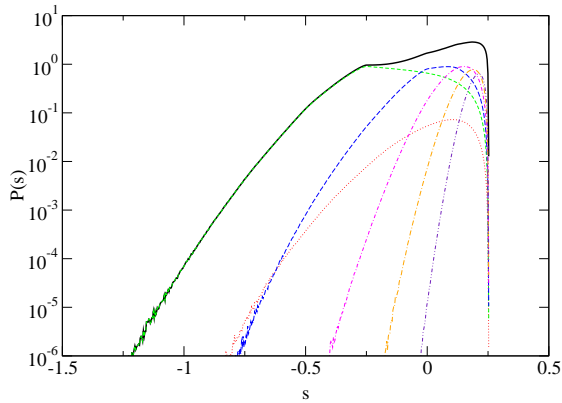}
\end{center}
\caption{\label{histo_fc} Distribution of depinning forces (bold line)
and contributions corresponding to growing distances along the front
between successive depinning sites. The larger the distance the
narrower the distribution and the closer from the critical threshold
$s^*$.}
\end{figure}

As developed above, the extremal dynamics gives a direct access to the
fluctuations of the driving force needed to depin the front site by
site. Most of these fluctuations simply correspond to the depinning of
near neighbors (which receive the largest contributions of the elastic
redistribution) and are highly sensitive to the details of the pinning
force disorder. The other part of these fluctuations corresponds to
depinning events taking place at a larger distance $d$ from the
previous depinning site. In other terms, the front can be regarded as
pinned over the distance $d$ between the two successive extremal
sites. Conditioning the depinning force distribution to this distance
$d$ between successive extremal sites thus allows to define
distributions of size dependent effective thresholds. (Note that in
the context of fracture, it can be seen as a distribution of effective
toughness\cite{CVHR-preprint03}). An effective threshold over a distance
$d$ can also be seen as the force needed to entirely depin a front of
size $d$. It can actually be shown \cite{SVR-IJMPC02,VSR-preprint03}
that as $d$ increases, the distributions $P(s;d)$ tend to peak and
approach the critical threshold $s^*$. This behavior is recovered in
our ultrametric variant as illustrated on Fig. \ref{histo_fc} where we
plotted the global distribution $P(s)$ and the contributions
$P(s;d_u)$ corresponding to the different ultrametric distances.

The typical elastic force fluctuations over a distance $d$ can
moreover be estimated: $\delta f^{el} \propto d^{-(1-\zeta)}$. Both
the width $\sigma_{s}(d)$ and the gap to the critical threshold
$\delta_{s}(d)=s^*-\langle s\rangle(d)$ of the center of these
distribution actually follow this scaling. As developed in
Ref. \cite{SVR-IJMPC02,VSR-preprint03} this property allows for a precise
extrapolation of the critical threshold by extrapolation of the linear
relationship between $\langle s\rangle(d)$ and $\sigma_{s}(d)$ up
to the force value canceling $\sigma_{s}$. Moreover this rescaling
results in a collapse of all distribution over a single master curve:

\be
P(s;d)=d^{1-\zeta} \chi \left[ d^{1-\zeta}(s^*-s) \right]\;.
\ee

\begin{figure}[t]
\begin{center}
\epsfig{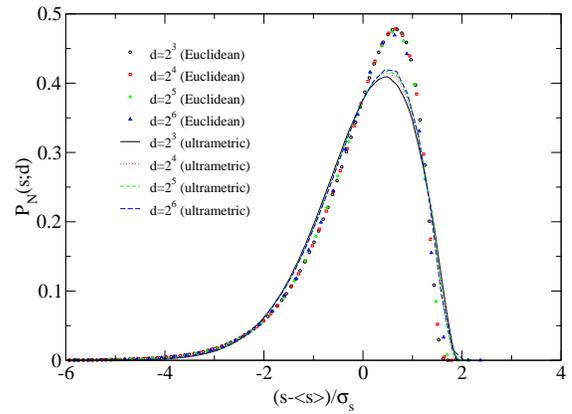}
\end{center}
\caption{\label{histo_fcR} After renormalization the centered unit
conditional distribution of depinning force fall onto a single curve
independently of the length $d$. The master curve obtained in the
ultrametric case (continuous lines) slightly differ from its Euclidean
counterpart (symbols).}
\end{figure}

We see on Fig. \ref{histo_fcR} that the choice of an ultrametric
distance slightly changes the shape of the distribution obtained after
rescaling. This may mean that these conditional distributions are more
sensitive to the boundary conditions and the loss of translation
invariance induced by the ultrametric model than a macroscopic
quantity such as the width of the interface.

\section{Conclusion}

We developed a depinning model based on the use of an ultrametric
distance. This choice allows to reduce the complexity of an elementary
step of computation to $\log_2 L$ with respect to the system size
$L$. We performed numerical simulations on systems of size $10^6$
more than two orders of magnitude larger than in previously published
results. We propose an estimate of the roughness exponent
$\zeta=0.390\pm0.01$ consistent with the values measured for the
Euclidean counterpart of the model. Moreover we obtained scaling
functions either identical or very close in both cases. Therefore the
choice of an ultrametric distance appears not to affect the
universality class of the depinning transition. 

Beyond the numerical efficiency, this model may thus serve as a
starting point for a real space renormalization analysis. Indeed a
coarse grained picture of the model preserves exactly the same
structure with ``dressed'' thresholds and front advances, and thus the
solution of a simple $L=2$ front model may open the way to an analytic
determination of critical exponents and universal scaling functions.

\begin{appendix}

\section[Appendix]{Numerical implementation of the model}

In this appendix we describe the algorithm used to run this
ultrametric model of depinning. As above mentioned  the structure of
the algorithm is roughly similar to the original one but steps {\bf B}
and {\bf E} are characterized by a complexity in $\log_2 L$ instead of
$L$.

Let us first consider the redistribution of elastic forces that take
place after a depinning event. The increment of force at a given site
$j$ only depends on its distance $d_u(i_0,j)=2^m-1$ to the extremal site
$i_0$. Actually $2^{m-1}$ different sites are at this same distance
from the extremal site $i_0$ and form a subtree (see
Fig. \ref{tree_force}). Instead of updating sequentially the elastic
force increment on every sites it is possible to update the elastic
force acting on the whole subtree. To this aim we define the force
$\varphi^{el}_{i,\ell}$ as the force transported by the $i^{th}$
branch of the $\ell^{th}$ level of the tree. The first level
($\ell=1$) here corresponds to the two branches at the root and the
level ($\ell=\log_2 L$) to the branches pointing toward the $L$ sites
of the front (the leaves of the tree). Using this definition the
elastic force acting on site $i$ is nothing but the sum of the forces
transported by its ancestor branches:

\be
\label{sumalongbranches}
f_i^{el} = \sum_{\ell=1}^{n=\log_2 L} \varphi^{el}_{\frac{i}{2^{n-\ell}},\ell}
\ee

In the hierarchical algorithm the step {\bf E} thus consists of
updating the branches of the force tree affected by the redistribution
of elastic forces :
$\varphi^{el}_{j_0,\ell}=\varphi^{el}_{j_0,\ell}+G_{i_0,m}\delta
h_{i_0}\;,\quad j_0=\{i_0/2^{n-\ell}\}$ where $\{ i\}$ is the
operation of exchanging the parity bit of $i$.

In the same spirit we can build a tree
to determine the value and location of the extremal site.
The determination of the extremal site requires {\it a priori} $L-1$
operations of elementary comparisons to find $s^*=\min_i (\gamma_i
-f_i^{el})$. After each depinning event, the elastic forces are
updated on every sites so that the same $L-1$ operations have to be
performed at each iteration. In the ultrametric model the situation
is slightly different because the elastic force on a site can be
written as a sum of the forces transmitted by the branches of the tree
(see Eq. \ref{sumalongbranches}). As a consequence two sites $i$ and
$j$ share all force components acting on branches above their nearest
common ancestor.

\bea
f_i^{el} &= &\Phi_m+
\sum_{\ell=m+1}^{n=\log_2 L} \varphi^{el}_{\frac{i}{2^{n-\ell}},\ell}\\
f_j^{el} &= &\Phi_m+
\sum_{\ell=m+1}^{n=\log_2 L} \varphi^{el}_{\frac{j}{2^{n-\ell}},\ell}\\
\Phi_m&=&\sum_{\ell=1}^{m} \varphi^{el}_{\frac{i}{2^{n-\ell}},\ell}
=\sum_{\ell=1}^{m} \varphi^{el}_{\frac{j}{2^{n-\ell}},\ell}
\eea

We can then perform the comparison operations level by level going up
the tree of elastic forces. To be more specific let us define the
following hierarchical structure. At level $n=\log_2 L$ we define

\be 
\sigma_{i,n}=\gamma_i\;,\quad \alpha_{i,n}=i
\ee

then we proceed iteratively up to the root of the tree:

\bea
\sigma_{i,p-1}&=&\min \left( \sigma_{2i,p}+\varphi^{el}_{2i,p},
\sigma_{2i+1,p}+\varphi^{el}_{2i+1,p} \right) 
\eea

\be
\begin{array}{lllll}
\alpha_{i,p-1}&=&\alpha_{2i,p} &\mathrm{if}
&\sigma_{2i,p}+\varphi^{el}_{2i,p} < \sigma_{2i+1,p}+\varphi^{el}_{2i+1,p} \\
&=&\alpha_{2i+1,p} &\mathrm{if} 
&\sigma_{2i,p}+\varphi^{el}_{2i,p} > \sigma_{2i+1,p}+\varphi^{el}_{2i+1,p} 
\end{array}
\ee

The location and the value of the extremal site are thus given by \be
s^*=\sigma_{1,0}\;,\quad i_0=\alpha_{1,0} \ee Without prior knowledge,
the computation for the determination of the extremal site of a given
conformation is simply the sum of all comparisons at each level of the
tree: $\sum_1^n2^{p-1}=2^n-1=L-1$. This is exactly the same result as
in the standard case. Consider now the situation after the depinning
event: the trap depth $t_{i0}$ is updated at the extremal site and
only $n=\log_2 L$ branches of the tree are altered by the elastic
force distribution. A sequence of $n$ comparisons thus allows to find
the new extremal site: we start at the former extremal site $i_0$,

\be
p=n\;: j_0=i_0\;,   \sigma_{j_0}= t_{j_0}\;,  \alpha_{j_0}=i_0
\ee

then for $p=n$ to $p=1$ we proceed iteratively to update the tree and
determine the value $\sigma_{1,0}$ and location $\alpha_{1,0}$ of the
new extremal site.

\be
 j_p=j_0/2^{n-p}\;,\quad k_p=\{j_p\}
\ee
 \be
\begin{array}{lll}
\sigma_{j_{p-1},p-1}&= &\min\left[
\sigma_{j_{p},p}+\varphi^{el}_{j_{p},p},
\sigma{k_{p},p}+\varphi^{el}_{k_{p},p} \right]
 \end{array}
\ee
\be
\begin{array}{lllll}
\alpha_{j_{p-1},p-1}&=&\alpha_{j_{p},p} &\mathrm{if} 
&\sigma_{j_{p-1},p-1}=\sigma_{j_{p},p}\\
&=&\alpha_{j_{p},p} &\mathrm{if} &\sigma_{j_{p-1},p-1}=\sigma_{k_{p},p}
\end{array}
\ee

\end{appendix}

\bibliography{vdb,depinning}

\end{document}